\documentclass[]{aastex631}

\usepackage{gensymb}
\usepackage{etoolbox}

\accepted{October 14, 2022}

\newcommand{\revision}{}

\submitjournal{AJ}

\shortauthors{Haqq-Misra and Fauchez}

\begin{document}

\title{Galactic settlement of low-mass stars as a resolution to the Fermi paradox}

\correspondingauthor{Jacob Haqq-Misra}
\email{jacob@bmsis.org}

\author[0000-0003-4346-2611]{Jacob Haqq-Misra}
\affiliation{Blue Marble Space Institute of Science, Seattle, WA, USA}


\author[0000-0002-5967-9631]{Thomas J. Fauchez}
\affiliation{American University, Washington DC, USA}


\begin{abstract}
An expanding civilization could rapidly spread through the galaxy, so the absence of extraterrestrial settlement in the solar system implies that such expansionist civilizations do not exist. This argument, often referred to as the Fermi paradox, typically assumes that expansion would proceed uniformly through the galaxy, but not all stellar types may be equally useful for a long-lived civilization. We suggest that low-mass stars, and K-dwarf stars in particular, would be ideal migration locations for civilizations that originate in a G-dwarf system. We use a modified form of the Drake Equation to show that expansion across all low-mass stars could be accomplished in 2\,\revision{Gyr}, which includes waiting time between expansion waves to allow for a close approach of a suitable destination star. This would require interstellar travel capabilities of no more than $\sim$\revision{0.3}\,ly to settle all M-dwarfs and $\sim$\revision{2}\,ly to settle all K-dwarfs. \revision{Even more rapid expansion could occur within 2\,Myr, with travel requirements of $\sim$10\,ly to settle all M-dwarfs and $\sim$50\,ly to settle all K-dwarfs.} The search for technosignatures in exoplanetary systems can help to place constraints on the presence of such a ``low-mass Galactic Club'' in the galaxy today.

\end{abstract}

\section{The Fermi Paradox} \label{sec:intro}

If the galaxy is teeming with intelligent life, then where is everybody? Various forms of this question have been posed by scientists interested in the search for extraterrestrial intelligence (SETI) \citep[e.g.,][]{hart1975explanation,jones1976colonization,tipler1980extraterrestrial}. Today the question is often described as the Fermi paradox \citep{gray2015fermi} or the Great Silence \citep{brin1983great}. The apparent absence of a galaxy-spanning extraterrestrial civilization or ``Galactic Club'' \citep{bracewell1976galactic}, so the argument goes, suggests that either interstellar travel is difficult or the evolution of life is a rare event. 

One of the most popular and widely cited versions of this argument was written by \citet{hart1975explanation}, who noted that expansion through the entire galaxy could be accomplished on timescales much shorter than the age of the galaxy:

\begin{quote}
    Assume that we eventually send expeditions to each of the 100 nearest stars. (These are all within 20 light-years of the Sun.) Each of these colonies has the potential of eventually sending out their own expeditions, and their colonies in turn can colonize, and so forth. If there were no pause between trips, the frontier of space exploration would then lie roughly on the surface of a sphere whose radius was increasing at a speed of $0.10c$. At that rate, most of our Galaxy would be traversed within 650,000 years. If we assume that the time between voyages is of the same order as the length of a single voyage, then the time needed to span the Galaxy will be roughly doubled. We see that if there were other advanced civilizations in our Galaxy they would have had ample time to reach us, unless they commenced space exploration less than 2 million years ago. \citep[][p.133]{hart1975explanation} 
\end{quote}

Numerous authors have attempted to refute this argument by providing alternate explanations for the apparent absence of a Galactic Club \citep[see e.g.,][]{webb2015if,cirkovic2018great,forgan2019solving}. This includes the possibility that extraterrestrial settlement of the solar system has already occurred but we have not discovered evidence of such visitation. For example, \citet{papagiannis1978we} suggested that \revision{signs} of extraterrestrial settlement might be discovered in the asteroid belt. Such possibilities cannot yet be dismissed. However, for the purpose of this paper, we will choose to accept the argument by \citet{hart1975explanation} at face value and will presume that no extraterrestrial expansion into the solar system has occurred. In this paper we also accept the assumption in the argument by \citet{hart1975explanation} and others \citep{jones1976colonization,tipler1980extraterrestrial} that extraterrestrial life will eventually develop technology and attempt expansion through the galaxy. But we disagree that the lack of extraterrestrial settlement of the solar system means that a Galactic Club does not exist. Instead, we suggest the hypothesis that interstellar expansion will primarily target low-mass systems in order to maximize the longevity of galactic settlement.

This hypothesis was recently suggested by \citet{hansen2021minimal} as a strategy for a technological civilization to respond to the evolution of its host star by migrating to a nearby lower-mass star. \citet{hansen2021minimal} noted that the barrier to interstellar travel depends on the radial distance at which a technological civilization can travel, but ``patience can reduce the distances required by two orders of magnitude relative to the long-term average distance between stars.'' We first assume that technological civilizations only arise on habitable planets orbiting G-dwarf stars (with $\sim$10\,Gyr main sequence lifetimes) because either biogenesis or complex life is more favored in such systems \citep[e.g.,][]{haqq2018we,haqq2019does}. This hypothesis then suggests that such civilizations would wait for an opportune moment to migrate to a nearby K- or M-dwarf star (with a $\sim$30-100+\,Gyr main sequence lifetime) during a close passage and before stellar evolution renders their home planet uninhabitable. \citet{hansen2021minimal} estimated that this motivation for civilizations to migrate from a short-lived to a long-lived star would imply that the fraction of extant civilizations around low-mass stars could range from 30\% to 72\% compared to G-dwarf stars. \citet{hansen2021minimal} noted the connection of their hypothesis with the argument by \citet{hart1975explanation}, but they explained that their focus was ``not how such civilizations may pursue expansion as a goal in and of itself.'' In this paper, we examine the implications of this hypothesis for interstellar expansion through the framework of the Drake equation  to suggest that a ``low-mass Galactic Club'' remains a possibility for our galaxy.

\section{The Drake Equation with Expansion}

The Drake equation \citep{drake1965radio} was originally developed by Frank Drake during a conference at the Green Bank Observatory in 1961. The purpose of the Drake equation is to show a functional relationship between the number of communicative civilizations is the galaxy today, $N$, and other factors that depend on the properties of planetary systems and life. A common expression of the Drake equation is:
\begin{equation}
    N=R_{*}\cdot f_{p}\cdot n_{e}\cdot f_{l}\cdot f_{i}\cdot f_{c}\cdot L,\label{eq:Drake}
\end{equation}
where $R_{*}$ is the rate of star formation, $f_{p}$ is the fraction of stars with planets, $n_{e}$ is the number of habitable planets per system, $f_{l}$ is the fraction of habitable planets that develop life, $f_{i}$ is the fraction of inhabited planets that develop intelligence, $f_{c}$ is the fraction of planets with intelligent life that develop communicative technology, and $L$ is the average communicative lifetime of technological civilizations. In this form, the Drake equation assumes that technological civilizations arise on a single planet and remain there for the duration of their detectable lifetimes.

A modification to the Drake equation that includes a factor to account for interstellar expansion was developed by \citet{walters1980interstellar}. The Drake equation with a factor for expansion, $E$, can be written as:
\begin{equation}
    N=R_{*}\cdot f_{p}\cdot n_{e}\cdot f_{l}\cdot f_{i}\cdot f_{c}\cdot L\cdot E,\label{eq:DrakeE}
\end{equation}
where
\begin{equation}
    E=1+f_{x}\sum_{n=1}\sum_{m=1}f_{s}^{n}ma_{m}\left[\left(m-1\right)a_{m}\right]^{n-1}.\label{eq:E}
\end{equation}
In Eq. (\ref{eq:E}), $f_{x}$ is the fraction of civilizations that wish to expand, $f_{s}$ is the fraction of stellar systems suitable for settlement, $a_{m}$ is the fraction of stars with $m$ neighboring stars within the radius $R_0$, and $n$ is the number of settlement waves. In the situation where no civilizations seek interstellar expansion, then $f_{x}=0$ and Eq. (\ref{eq:DrakeE}) reduces to the original Drake equation. When $f_{x}>0$, then the factor $E$ causes an increase in the value of $N$ depending on the number of available sites for each inhabited system and the number of settlement waves. 

The extent to which a civilization can easily traverse the galaxy depends on the number of stars within its navigable radius. A civilization with limited interstellar travel capabilities would have a small value of $R_0$ and may find no destinations that it can reach. If this is typical for all civilizations, then $a_{m=1}=1$ and $a_{m\ne1}=0$, which reduces Eq. (\ref{eq:DrakeE}) to the original Drake equation. (Note that $a_{m=1}=1$ in this case because a star is considered to be its own neighbor in Eq. (\ref{eq:E}).) A civilization with a larger value of $R_0$ may find one or more stars on average that could be a destination for settlement. \citet{walters1980interstellar} showed that Eq. (\ref{eq:E}) diverges when
\begin{equation}
    f_{s}\sum_{m=1}\left(m-1\right)a_{m}>1,\label{eq:diverge}
\end{equation}
and $n$ is limitless, which indicates that galactic-scale settlement is possible. If the left-side of Eq. (\ref{eq:diverge}) is less than 1, then Eq. (\ref{eq:E}) converges, which indicates that any settlement is limited. 

We first consider the case in which all civilizations have only one destination star within $R_0$, so that $a_{m=2}=1$ and $a_{m\ne2}=0$. In this case, the condition in Eq. (\ref{eq:diverge}) is only satisfied when $f_s>1$, but $f_s$ by definition is less than or equal to 1. This case can lead to some interstellar migration but cannot satisfy the condition in the scenario by \citet{hart1975explanation} for rapid galactic expansion.

Next we consider the case in which all civilizations have two destination stars within $R_0$, so that $a_{m=3}=1$ and $a_{m\ne3}=0$. In this case, the condition in Eq. (\ref{eq:diverge}) is satisfied when $f_s > 0.5$. A value of $f_s$ this large could be possible if civilizations are able to adapt to a wide range of stellar systems. The analysis by \citet{walters1980interstellar} assumed that suitable stars for settlement ``are ones which have environments nearly identical to that of the home planet and can be immediately exploited'' such as a ``breathable atmosphere,'' which the authors estimate to give $f_s \sim 0.07$. But such constraints on the destination are unnecessary, as \revision{settlers could} construct artificial megastructures \revision{that} would allow \revision{them} ``to set-up-shop in orbit about the new star without depending on a habitable planet'' \citep{jones1976colonization}. If most or all stars within $R_0$ are suitable for settlement, then galactic-scale expansion can occur quickly.

The timescale for galactic expansion depends on the length of time that a civilization must wait before attempting interstellar migration. \citet{hansen2021minimal} developed an expression for the equivalent time between close stellar passages, $\tau$, as:
\begin{equation}
    \tau=\frac{1}{\rho V\pi R_{0}^{2}},\label{eq:tau}
\end{equation}
where $\rho$ is the local stellar density and $V$ is the average velocity of encounter. \citet{hansen2021minimal} estimate $n=0.1$\,pc$^{-3}$ and $V=48$\,km\,s$^{-1}$ for the solar neighborhood and note that the values of $\rho$ and $V$ can be higher or lower \revision{in} different regions of the galaxy. For a civilization in the solar neighborhood with $R_0=1,000$\,au ($\sim$0.02\,ly), close stellar encounters would occur once every 2.8\,Gyr. The galaxy is about 9\,Gyr old, so only $n=3$ waves of settlement at this slow pace could have occurred so far. 

Rapid galactic expansion thus requires a larger average $R_0$ for civilizations, such as the $20$\,light-year ($\sim$10$^6$\,au) radius used by \citet{hart1975explanation}. Returning to the example with $a_{m=3}=1$ and $a_{m\ne3}=0$, where all civilizations have two destination stars, Eq. (\ref{eq:E}) requires at least $n=35$ waves in order for migration to spread across 10$^{11}$ stellar systems, assuming $f_s = 1.0$. If these waves are spread evenly across the first 4.5\,Gyr of the galaxy's history, with half the time allocated to travel and the other half to waiting for a close stellar approach, then each wave would occur every 64\,Myr and would require a minimum travel distance of $R_0 \ge 9,300$\,au (0.15\,ly). If instead we choose $f_s = 0.51$, and we assume that only $f_s\times10^{11}$\,stars are settled, then this would \revision{need} $n=1027$ waves to expand across the galaxy, which would occur every 2.2\,Myr and require $R_0 \ge 50,000$\,au (0.79\,ly). This value of $R_0$ is about 20\% the distance from Earth to Proxima Cen today. We note that this calculation assumes solar neighborhood values for $\rho$ and $V$ in Eq. (\ref{eq:tau}) and also presumes that each close stellar encounter will provide new (unsettled) destinations for the expanding civilization that do not overlap with previously settled systems. Whereas \citet{hart1975explanation} and others based their calculations on a galaxy with fixed interstellar distances, the possibility of waiting for close stellar approaches allows an expanding civilization to capitalize on the motion of stars in the galaxy to reduce the total travel distance.

We can consider five settlement scenarios on a shorter time period of 2.0\,Gyr, again with half of this time allocated to travel and the other half to waiting, summarized in Table \ref{tab:1}. Each scenario is defined by the suitable star fraction ($f_s$), which we use in to find the minimum number of neighbor stars ($M$, so that $a_{m=M}=1$ and $a_{m\ne M}=0$) needed to satisfy Eq. (\ref{eq:diverge}). We solve Eq. (\ref{eq:E}) for the number of settlement waves ($n$) by assuming $E = f_s\times10^{11}$, and we calculate the average travel time for each wave ($t$) as the total time period divided by $2n$. We calculate the average wait time for a close stellar passage ($\tau$) as the total time period divided by $2n(M-1)$, and we then use this value of $\tau$ in Eq. (\ref{eq:tau}) to calculate the minimum travel distance ($R_0$). The five scenarios show that settlement of the full galaxy ($f_s=1.0$), all M-dwarf stars ($f_s=0.73$), half the galaxy ($f_s=0.51$), all K-dwarf stars ($f_s=0.13$), and all G-dwarf stars ($f_s=0.06$) could occur within 2.0\,Gyr without the need for interstellar travel beyond a few light years. The allocation for travel time is large in these scenarios, but any time not needed for migration could then extend the time needed to wait for a suitable star to approach.

\begin{table}[ht!]
\centering
\caption{Five scenarios for the settlement of $f_s\times10^{11}$\,stars in 2.0\,Gyr are defined by the suitable star fraction ($f_s$) and the minimum number of neighbor stars ($M$), which give the number of waves ($n$), average travel time for each wave ($t$), the average wait time for a close stellar passage ($\tau$), and the minimum travel distance ($R_0$). The 2.0\,Gyr time period is divided with half allocated to travel time and half to waiting.\label{tab:1}}
\begin{tabular}{ccc|cccc} 
Settlement scenario & $f_s$ & $M$ & $n$ & $t$ & $\tau$ & $R_0$ \\
\hline\hline
Full galaxy         & 1.0   & 3  & 35   & 28.6\,Myr  & 14.3\,Myr  & 1.39$\times10^4$\,au (0.22\,ly) \\ 
All M-dwarf stars   & 0.73  & 3  & 62   & 16.1\,Myr  & 8.06\,Myr  & 1.85$\times10^4$\,au (0.29\,ly)  \\
Half galaxy         & 0.51  & 3  & 1027 & 0.974\,Myr & 0.487\,Myr & 7.53$\times10^4$\,au (1.2\,ly)  \\
All K-dwarf stars   & 0.13  & 9  & 508  & 1.97\,Myr  & 0.246\,Myr & 1.06$\times10^5$\,au (1.7\,ly)  \\
All G-dwarf stars   & 0.06  & 18 & 936  & 1.07\,Myr  & 0.0628\,Myr & 2.10$\times10^5$\,au (3.3\,ly)  \\
\hline                       
\end{tabular}
\end{table}

\begin{table}[ht!]
\centering
\caption{Same as Table \ref{tab:1}, but the five scenarios are now calculated for the settlement of $f_s\times10^{11}$\,stars in 2.0\,Myr, to be comparable with the argument by \citet{hart1975explanation}. The 2.0\,Myr time period is divided with half allocated to travel time and half to waiting.\label{tab:2}}
\begin{tabular}{ccc|cccc} 
Settlement scenario & $f_s$ & $M$ & $n$ & $t$ & $\tau$ & $R_0$  \\
\hline\hline
Full galaxy         & 1.0   & 3  & 35   & 28.6\,kyr  & 14.3\,kyr  & 4.39$\times10^5$\,au (6.9\,ly)  \\ 
All M-dwarf stars   & 0.73  & 3  & 62   & 16.1\,kyr  & 8.06\,kyr  & 5.85$\times10^5$\,au (9.2\,ly)  \\
Half galaxy         & 0.51  & 3  & 1027 & 0.974\,kyr & 0.487\,kyr & 2.38$\times10^6$\,au (38\,ly)  \\
All K-dwarf stars   & 0.13  & 9  & 508  & 1.97\,kyr  & 0.246\,kyr & 3.35$\times10^6$\,au (53\,ly)  \\
All G-dwarf stars   & 0.06  & 18 & 936  & 1.07\,kyr  & 0.0628\,kyr & 6.63$\times10^6$\,au (100\,ly)  \\
\hline                       
\end{tabular}
\end{table}

We examine these same five scenarios but with a much shorter time period of 2\,Myr in order to make a more direct comparison with the assumptions in the argument by \citet{hart1975explanation}. These results are summarized in Table \ref{tab:2}. The number of expansion waves remains the same for each scenario, and the timescales $t$ and $\tau$ have decreased into the thousands of years. The required interstellar travel distance for these scenarios requires no more than $R_0 = 10$\,ly to settle all M-dwarf systems and $R_0 = 53$\,ly to settle all K-dwarf systems. A scenario involving the settlement of all G-dwarf systems could be accomplished with $R_0 = 100$\,ly. However, this scenario, as well as full galaxy settlement, can be excluded based on our assumption that the solar system has not been settled. But the lack of extraterrestrial settlement of the solar system remains consistent with expansion that is limited to half the galaxy, M-dwarf stars, or K-dwarf stars.

This exercise demonstrates that the argument put forth by \citet{hart1975explanation} is made only more poignant by showing that civilizations can leverage close stellar encounters to rapidly expand across the galaxy, \revision{without the need for relativistic spaceflight}. This behavior has been demonstrated by others using more sophisticated models, which also show that the existence of a Galactic Club cannot necessarily be excluded \citep[e.g.,][]{carroll2019fermi}. In particular, we note that a low-mass Galactic Club, originating from a parent G-dwarf system, would have had plenty of time to develop in the history of the galaxy without us taking any notice of its activities.

\section{The K-dwarf Galactic Club}

All attempts to explain the problem posed by \citet{hart1975explanation} must grapple with the unknown characteristics of extraterrestrial sociology. We do not know if extraterrestrial civilization exists, and if it does, we have no knowledge of their motives. Several authors have noted that exponential population growth through the galaxy is unsustainable \citep{von1975population,newman1981galactic,haqq2009sustainability,mullan2019population}, so any attempt at galactic-scale expansion must be for reasons other than to satisfy the demands of a growing population or increasing energy consumption. But we do not know much more about whether or not such galactic-scale expansion would be commonplace or desirable for technological civilizations in general. \citet{newman1981galactic} performed a detailed mathematical analysis of population diffusion in the galaxy and concluded that only long-lived civilizations could have established a Galactic Club; however, such a possibility was excluded because the authors ``believe that their motivations for colonization may have altered utterly.'' Although it remains possible that long-lived technological civilizations do not expand, it also remains possible that such civilizations pursue galactic settlement in order to ensure their longevity. The numerical simulations by \citet{carroll2019fermi} found solutions where ``our current circumstances are consistent with an otherwise settled, steady-state galaxy.'' But why would an  extraterrestrial civilization expand across the galaxy but not settle the solar system? We suggest, following the hypothesis of \citet{hansen2021minimal}, that an expanding civilization will preferentially settle on low-mass K- or M-dwarf systems, avoiding higher-mass stars, in order to maximize their longevity in the galaxy.

We further suggest that K-dwarf stars would be ideally suited as a migration destination for a civilization that arises in a G-dwarf system. As mentioned earlier, \cite{walters1980interstellar} hypothesized that stars suitable for settlement should provide ``environments nearly identical to that of the home planet.'' If we expand our consideration of a suitable environment to the star-planet system, then K-dwarf systems would provide \revision{a} suitable option as the destination for a civilization accustomed to life around a G-dwarf star. K-dwarfs represent about 13$\%$ of the galaxy's stellar demographics (compared to only 6$\%$ for G-dwarfs and 73$\%$ for M-dwarfs) and have a life expectancy ranging from 17 Gy to 70 Gy for the highest and lowest mass K-dwarfs, respectively, compared to the 10 Gy life expectancy of the sun. \cite{Arney2019} noted that K-dwarf spectra share far more similarities with G-star spectra compared to M-dwarfs. Planets can be also found in the habitable zone of such systems \citep{kopparapu2013habitable} without being tidally locked, which would generally have greater similarities to G-dwarf habitable zone planets compared to planets in an M-dwarf system. Furthermore, stellar activity, which could be detrimental to life as-we-know-it, is significantly weaker on K-dwarfs than M-dwarfs. This perspective is supported by the Exoplanet Science Strategy report
\citep{NASM2018}, which stated that to maximize our chances of discovering habitable worlds and life elsewhere, we must also seek observations of temperate terrestrial planets orbiting ``Sun-like stars'' (i.e., F, G, and K types). K-dwarfs may be at the sweet spot for both biosignature detection and extraterrestrial expansion, as they offer a larger population than G-dwarfs, a much longer lifetime, and a smoother environmental transition compared to M dwarfs systems.

\revision{Galactic-scale settlement is not the only reason that could motivate an extraterrestrial civilization to visit nearby star systems. For example, \citet{2002Mercu..31e..14Z} suggested that G-dwarf systems with inhabited planets, such as Earth, may provide attractive targets for extraterrestrial visitors who are interested in studying the evolution of life on other worlds. This form of exploration could also involve remote exploration, such as the use of self-replicating probes \citep[e.g.,][]{cotta2009computational,nicholson2013slingshot,ellery2022self}. The analysis by \citet{walters1980interstellar} even assumed that that the most attractive sites for interstellar settlement would be those that already host life. Such possibilities cannot be dismissed and suggest that the effort to search for extraterrestrial artifacts \citep[e.g.,][]{bracewell1976galactic,freitas1983search,tough1998small,haqq2012likelihood,benford2019looking,benford2021drake} is worth pursuing. However, if strong constraints can be placed on the presence of extraterrestrial technology within the solar system, then one possible explanation is that extraterrestrial travelers prefer to visit other types of stellar systems.}

Extraterrestrial expansion through the galaxy remains viable. We can exclude scenarios in which all G-dwarf stars would have been settled by now, but the possibility remains open that a Galactic Club exists across all K-dwarf or M-dwarf stars. The search for technosignatures in low-mass systems provides one way to constrain the presence of such a Galactic Club \citep[e.g.,][]{lingam2021life,socas2021concepts,wright2022case,haqq2022searching}. Existing searches to-date have placed some limits on radio transmissions \citep[e.g.,][]{harp2016seti,enriquez2017breakthrough,price2020breakthrough,zhang2020first} \revision{and optical signals} \citep[e.g.,][]{howard2007initial,tellis2015search,schuetz2016optical} that might be associated with technological activity, but such limits can only weakly constrain the Galactic Club hypothesis. Further research into understanding the breadth of possibilities for detecting extraterrestrial technology will become increasingly important as observing facilities become more adept at characterizing terrestrial planets in low-mass exoplanetary systems.

K-dwarf systems may be ideal targets to search for biosignatures and technosignatures, but such planets can be \revision{difficult} to detect. Many K-dwarf stars are too big to allow atmospheric characterization of planets with transit observations (and even planet detection can be difficult), while M-dwarf systems can more easily be characterized with transits. The planet-star separation in K-dwarf systems is also too small for many direct imaging missions, whereas G-dwarf systems are better targets for direct imaging. \revision{K-dwarf systems may be} suitable sites for migration, \revision{but placing observational constraints on the presence of a K-dwarf Galactic Club may be particularly challenging.}

\begin{acknowledgments}
J.H.M. gratefully acknowledges support from the NASA Exobiology program under grant 80NSSC22K1009. The authors also acknowledge support from the Goddard Space Flight Center (GSFC) Sellers Exoplanet Environments Collaboration (SEEC), which is supported by the NASA Planetary Science Division's Research Program. Any opinions, findings, and conclusions or recommendations expressed in this material are those of the authors and do not necessarily reflect the views of their employers or NASA.
\end{acknowledgments}


\bibliography{main}{}
\bibliographystyle{aasjournal}

\end{document}